\documentclass[aps,pra,twocolumn,groupedaddress]{revtex4}

\usepackage{graphics}% Include figure files
\usepackage{graphicx}% Include figure files
\usepackage{dcolumn}% Align table columns on decimal point
\usepackage{bm}% bold math

\newcommand{\de}{\partial}

\newcommand{\eq}[2]{\begin{equation} \label{#1} #2 \end{equation}}
\newcommand{\eps}{\epsilon}

\newcommand{\etal}{{\em et al.}}

\begin{document}

\title{Microcavity polariton-like dispersion doublet in resonant Bragg gratings}

\author{Fabio Biancalana\textsuperscript{1},
Leonidas Mouchliadis\textsuperscript{2}, Celestino Creatore\textsuperscript{3}, Simon Osborne\textsuperscript{4} and Wolfgang Langbein\textsuperscript{2}}

\affiliation{\textsuperscript{1}Max Planck Institute for the Science of Light, G\"{u}nther-Scharowsky-Str. 1/Bau 24, 91058 Erlangen, Germany\\ \textsuperscript{2}School of Physics and Astronomy,
Cardiff University, The Parade, CF24 3AA, Cardiff, UK \\ \textsuperscript{3}Department of Physics ``A. Volta", Universit\`{a} degli Studi di Pavia, via Bassi 6, Pavia, Italy \\ \textsuperscript{4}Tyndall National Institute, Lee Maltings, Prospect Row, Cork, Ireland}

\date{\today}

\begin{abstract}
Periodic structures resonantly coupled to excitonic media allow the existence of extra intragap modes
('Braggoritons'), due to the coupling between Bragg photon modes and 3D bulk excitons. This induces unique and unexplored dispersive
features, which can be tailored by properly designing the photonic bandgap around the exciton resonance.
We report that 1D Braggoritons realized with semiconductor gratings have the ability to mimic
the dispersion of quantum-well microcavity polaritons. This will allow the observation of new nonlinear phenomena, such as slow-light-enhanced
nonlinear propagation and an efficient parametric scattering at two 'magic frequencies'.
\end{abstract}

\maketitle

Since the pioneering proposals by Yablonovitch and John \cite{yablo}, photonic crystals (PhCrs), structures characterized by a spatially periodic dielectric function, have attracted enormous attention due to their rich physics. In particular, the occurrence of photonic bandgaps (PBGs), i.e. frequency regions where the propagation of light is strongly inhibited and the extraordinary ability to manipulate and control the photonic flow, make PhCrs very appealing for many applications \cite{yablo}. Furthermore, the current accurate engineering of photonic states allows to investigate light-matter phenomena in PhCrs; one of the most fascinating being the strong-coupling regime of light-matter interaction \cite{strongcoupling}.
Non-trivial modifications of the photonic dispersion are expected to occur when PBG materials are coupled to polarizable excitonic media \cite{vardeny,grundmann}, as in this case the true eigenmodes of the mixed photonic-excitonic system are exciton-polaritons \cite{excitons}, i.e. the normal modes born from the strong coupling between Wannier-Mott (WM) excitons \cite{vardeny,grundmann} and Bragg photons propagating in the PBG structure.

In this Letter, we show through theoretical analysis that, by embedding a three-dimensional (3D) exciton resonance within the PBG of a one-dimensional (1D) Bragg grating, the dispersion relation of the resulting exciton-polariton states can, under certain conditions, mimic the dispersion of a {\em doublet} of microcavity (MC) polaritons, i.e. the quasiparticles resulting from the coupling between two-dimensional (2D) excitons in a quantum well and the optically confined photons in a semiconductor planar microcavity \cite{excitons}. The very peculiar dispersion features of these Bragg polariton modes have not been unveiled previously, and, as we demonstrate here, give rise to a wide variety of new nonlinear phenomena, absent in standard gratings, such as slow-light-enhanced nonlinear propagation and an ultra-efficient parametric scattering at two 'magic frequencies'. In addition, owing to the extremely small effective mass of these novel hybrid exciton-photon modes, new routes for the appearance of macroscopic coherence in solid state systems are opened.

It is possible to rigorously show, starting from Maxwell's equations, that the interaction between the electromagnetic field in a periodic medium (like the 1D grating here examined) and the exciton-polarization wave, is described (in dimensionless variables) by the following {\em polaritonic coupled-mode equations} (PCMEs):
\begin{widetext}
\begin{eqnarray}
&&i[\de_{\tau}+\tilde{\gamma}_{\rm ph}]f+i\de_{\zeta}f+\kappa
b+p_{f}/2=0, \label{m1}\\
&&i[\de_{\tau}+\tilde{\gamma}_{\rm ph}]b-i\de_{\zeta}b+\kappa
f+p_{b}/2=0, \label{m2}\\
&&i\de_{\tau}p_{f}+(i\tilde{\gamma}_{\rm x}+d)p_{f}+\left[|p_{f}|^{2}+2|p_{b}|^{2}
\right]p_{f}+\rho f=0, \label{m3}\\ 
&&i\de_{\tau}p_{b}+(i\tilde{\gamma}_{\rm x}+d)p_{b}+\left[|p_{b}|^{2}+2|p_{f}|^{2} \right]p_{b}+\rho b=0,
\label{m4}
\end{eqnarray}\end{widetext} In Eqs. (\ref{m1}-\ref{m4}), the fields $\{f,b\}$ represent the amplitude of the slowly-varying envelopes of the forward and backward propagating electric field respectively,
while $\{p_{f},p_{b}\}$ are the corresponding quantities for the exciton polarization field. Spatial dispersion terms have been omitted for simplicity, as their contribution to the propagation dynamics is negligible.  The dimensionless temporal and longitudinal spatial variables are $\tau\equiv t/t_{0}$ and $\zeta\equiv z/z_{0}$ respectively, with $t_{0}\equiv\eps_{b}/\omega_{0}$ and $z_{0}\equiv ct_{0}/\sqrt{\eps_{b}}$,
$\eps_{b}$ being the average background dielectric function of the grating, and $\omega_{0}$ is the central pulse frequency. The 1D grating is described by the dielectric function $\eps(z)=\eps_{b}\{1+\mu[e^{ik_{B}z}+e^{-ik_{B}z}]/2\}$, where $\mu\ll 1$ is the depth of the grating, and $k_{B}\equiv 2\sqrt{\eps_{b}}\omega_{0}/c$ is the grating Bragg wavenumber.
The dimensionless frequency detuning between the Bragg frequency $\omega_{B}$ and the exciton resonant frequency $\omega_{\rm x}$ is given by $d\equiv\Delta\omega t_{0}$,
where $\Delta\omega=\omega_{B}-\omega_{\rm x}$. The linear absorption of the background medium $\gamma_{\rm ph}$ and exciton oscillator damping $\gamma_{\rm x}$ (i.e.
the exciton homogeneous linewidth) are also considered in our model through the dimensionless quantities $\tilde{\gamma}_{\rm ph,x}\equiv t_{0}\gamma_{\rm ph,x}$.
The polariton splitting $\omega_{c}$, which measures the interaction between the transverse components of the excitonic polarization and the retarded electromagnetic field,
is specified by the dimensionless parameter $\rho\equiv t_{0}\omega_{c}^{2}/(2\omega_{\rm x})$. Finally, solution of the linearized Eqs. (\ref{m1}-\ref{m4}) yields the dispersion
relation of Bragg polaritons $\delta=\delta(q)$, where $\delta\equiv\Delta\omega/\omega_{0}$ and $q\equiv\Delta k/k_{0}$ are the frequency and wavenumber detunings from the
bandgap center respectively, and have been introduced via a phase shift of the fields of the kind $\exp(iq\zeta-i\delta\tau)$.

% design
Throughout the whole paper we use a design based on Zinc Oxide (ZnO) as a representative example of our theoretical calculations.
ZnO has received substantial attention in recent years as a material for blue light emitting devices, transparent electrodes, and solar cells. This is due to the robustness of its exciton resonances, which remain stable up to room temperature, and its exceptionally large binding energy and oscillator strength \cite{medard}. Following recent advances in ZnO growth and deposition, clean exciton resonances have been produced using molecular beam epitaxy \cite{ZnO},
pulsed-laser deposition \cite{grundmann}, sputtering \cite{chichibu} and led to the realization of a ZnO MC \cite{grundmann}. We use ZnO as our excitonic material [see also Fig. \ref{fig3n}(e)], and we therefore design the photonic crystal in order to have the PBG located around $\omega_{\rm x}$. The second kind of layer is made of ZrO$_{2}$. The free FX$_{A}^{n=1}(\Gamma_{5})$-exciton binding energy of ZnO is $\hbar\omega_{c}\approx 60$ meV, and the exciton central frequency is $\hbar\omega_{\rm x}\approx 3.3771$ eV, which corresponds to $\lambda_{\rm x}\simeq 367.4$ nm \cite{ZnO}.
The linear absorption coefficient and the exciton homogeneous linewidth of ZnO depend considerably on the fabrication technique. However, at  $T=5$ K we make the estimates $\gamma_{\rm ph}\approx 2.5$ meV and $\gamma_{\rm x}\approx 0.25$ meV \cite{hazu}. The strength of the exciton-photon coupling is thus $\rho/t_{0}=\omega_{c}^{2}/(2\omega_{\rm x})\simeq 5.33\times
10^{-4}$ eV. Near the exciton transition frequency, ZnO has a refractive
index $n_{1}\simeq 2.37$, while for ZrO$_{2}$
$n_{2}\simeq 2.326$ at the same wavelength. Thus we have the following grating
parameters, to be used in Eqs. (\ref{m1}-\ref{m4}): $\mu=\Delta
n^{2}/\eps_{b}\simeq 0.0187$, and
$\kappa=\mu\eps_{b}/4\simeq 0.0258\ll 1$,
satisfying the shallow grating condition on which our PCMEs are based.
The quarter wavelength condition gives the layers' width: $L_{1}=38.8$ nm and $L_{2}=39.9$ nm for ZnO and ZrO$_{2}$ respectively.
The nonlinear refractive index of ZnO at $\lambda_{\rm x}$ is estimated to be $n_{\rm NL}({\rm ZnO})\approx 10^{-11}$ m$^{2}$/W at $5$ $^{\circ}$K, about 9 orders of magnitude larger than that of bulk silica \cite{zhang}.

% Dispersion
Equations (\ref{m1}-\ref{m4}) represent the foundational result of
this Letter. Previous attempts to describe polaritonic gratings have
been made in the framework of the linear transfer matrix method (TMM) \cite{vardeny},
approaches that do not take into account the large nonlinear optical response
of excitons near resonance, or by using a model based on Maxwell-Bloch equations \cite{previous}, but without taking into account the crucial importance of both forward and backward propagating
exciton polarization waves. Our approach provides a simple, complete and tractable set of
equations that avoids the complications of the TMM, but at the same time yields results
that are very close to the ones obtained by numerically solving the full Maxwell equations coupled to the material equations.

\begin{figure}
\includegraphics[width=8cm]{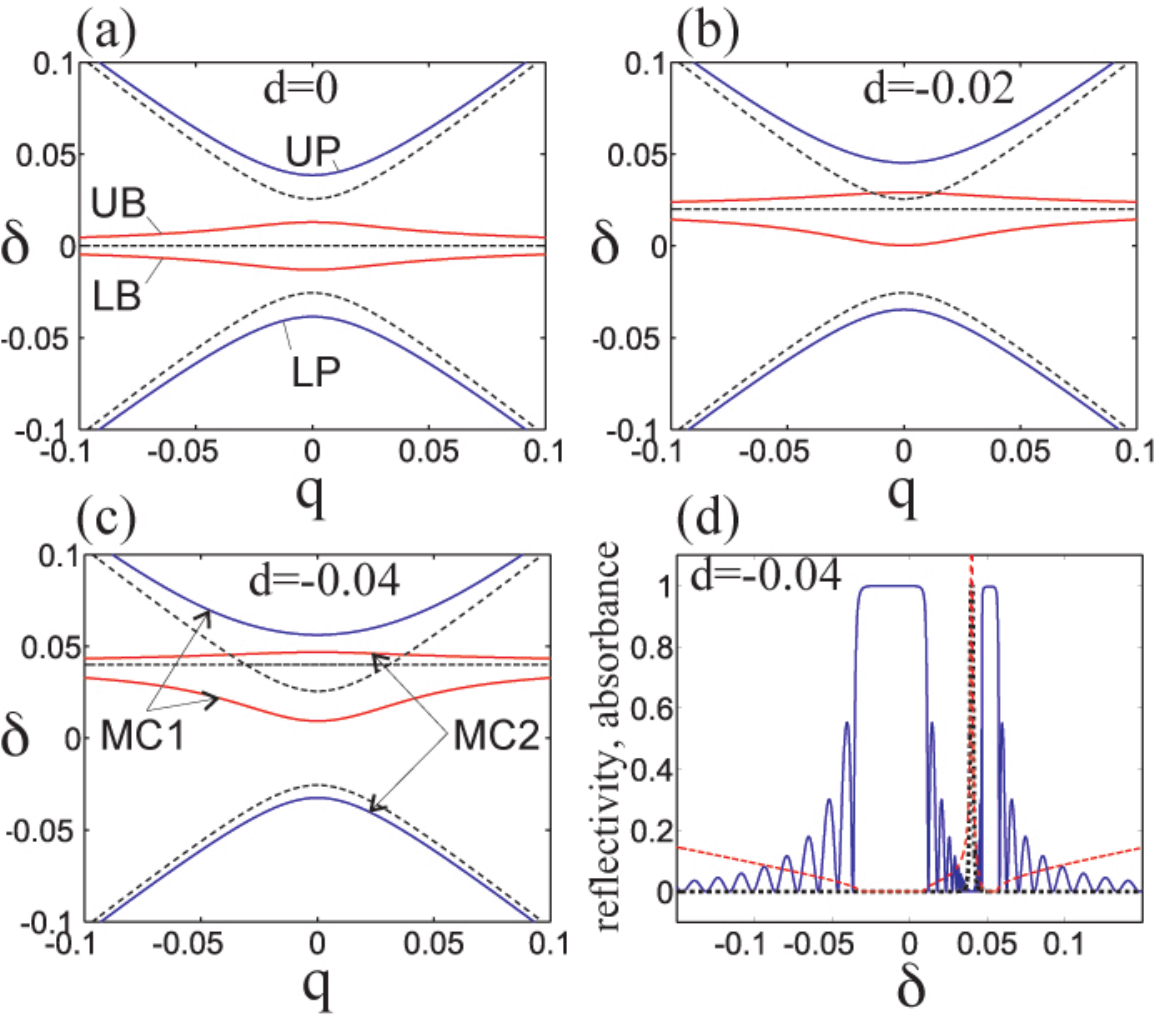}
\caption{\small (Color online) (a,b,c) Polariton dispersion on the $(\delta,q)$-plane for three different values of $d$.
Dashed lines are the bare Bragg-photon and exciton dispersions. Blue solid lines indicate upper-photon (UP) and lower-photon (LP) branches, red solid
lines are upper-Braggoriton (UB) and lower-Braggoriton (LB) branches. (d) Reflectivity $R_{j}$ (blue solid line) and absorbance $A_{j}$ (black dotted line)
as functions of $\delta$ for a finite grating ($L\simeq 14$ $\mu$m for our proposed ZnO design).
The dispersion is shown with red dashed lines. \label{fig1n}}
\end{figure}

Figs. \ref{fig1n}(a-c) show the evolution of the coupled and bare {\em linear}
dispersions when varying $d$ from vanishing to negative
values. Coherent coupling
between 3D bulk excitons and Bragg photons gives rise to two new intragap modes (Bragg polaritons, or {\em Braggoritons} \cite{vardeny}), the dispersion of
which can be efficiently tailored by simply tuning $d$, for instance by slightly
modifying the grating parameters around the fixed exciton resonant
frequency.

\begin{figure}
\includegraphics[width=8cm]{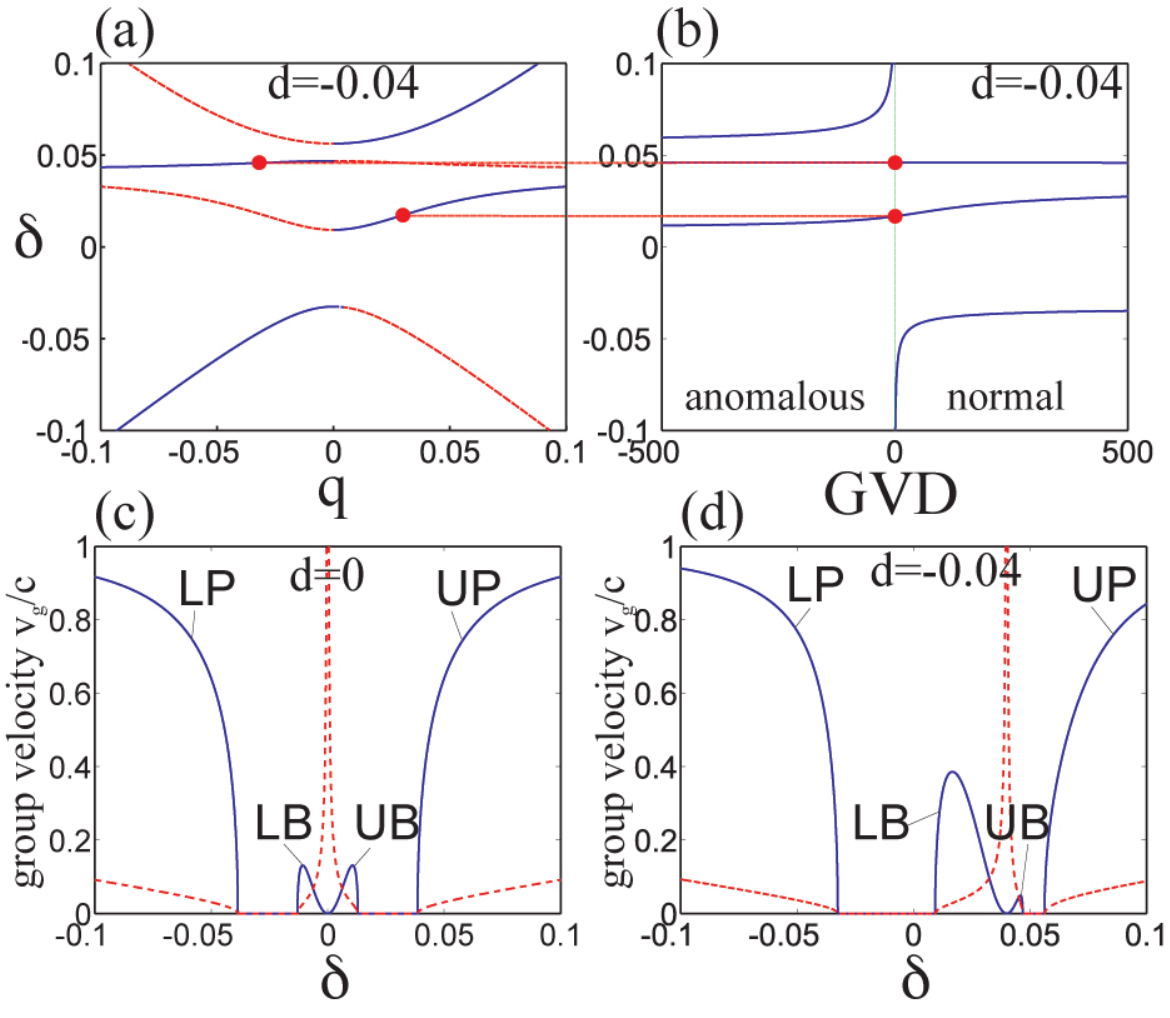}
\caption{\small (Color online) (a) Physical parts of the dispersion branches (for which $R(\delta)\le 1$) are indicated with blue solid lines, for $d=-0.04$. Red dots indicate the position of the two inflection points on the UB and LB branches, which are located at $q$'s that have the same magnitude but opposite signs, see also Fig. \ref{fig3n}(b). (b) GVD ($\de^{2}q_{j}/\de\delta^{2}$) as a function of $\delta$. Red dots indicate zero-GVD points. (c,d) Solid blue lines are the group velocities $v_{g,j}/c=(\de q_{j}/\de\delta)^{-1}$ as functions of $q$, for $d=0$ and $d=-0.04$ respectively. Red dashed lines indicate the dispersion. \label{fig2n}}
\end{figure}

\begin{figure}
\includegraphics[width=8cm]{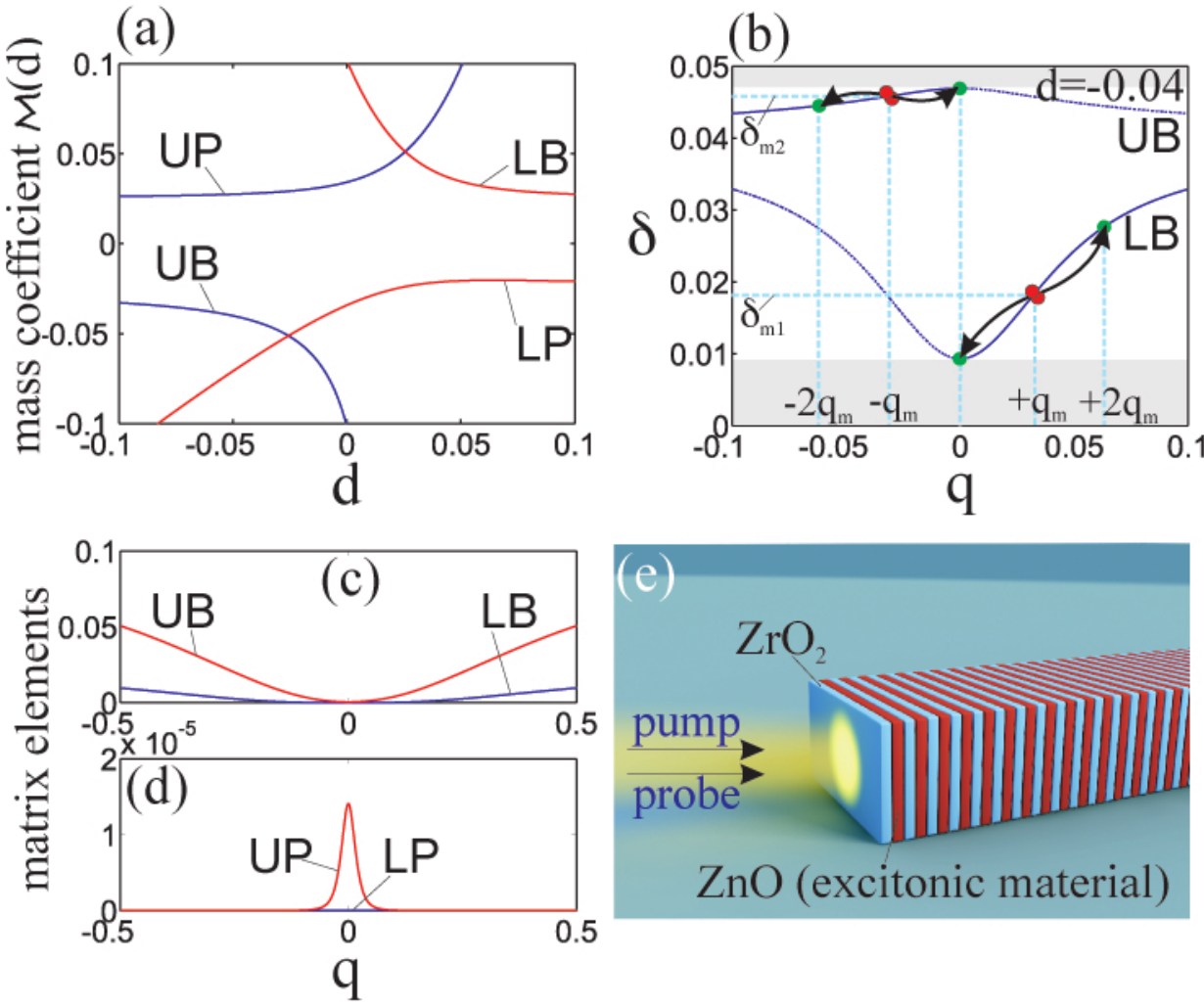}
\caption{\small (Color online)  (a) Effective polariton mass parameter $\mathcal{M}_{j}$ as a function of $d$. (b) Schematics of the parametric amplification process occurring in the proximity of the inflection points of both Braggoriton branches, for $d=-0.04$. Grey areas indicate the photonic sub-gaps.
(c) Matrix element for the intraband parametric scatterings shown in (b), for the LB branch (blue line) and for the UB branch (red line). (d) Same as (c) but for the LP branch (blue line)
and for the UP branch (red line). (e) Sketch of our proposed design for a polaritonic Bragg stack. The excitonic material is ZnO (shown in red, $\lambda_{\rm x}=367.4$ nm), while the second material is ZrO$_{2}$ (shown in blue). \label{fig3n}}
\end{figure}

The four branches in question [which we label upper- and lower-photon (UP,LP) and upper- and lower-Braggoriton (UB,LB) branches,
see Figs. \ref{fig1n}(a-c)] have a striking resemblance with the dispersions of a {\em doublet} of coupled MC polaritons \cite{kavokin}.
As shown in Fig. \ref{fig1n}(c) for a detuning $d=-0.04$, the pair (UP,LB) form the first half of the doublet (named MC1), while its counterpart (named MC2), formed by the pair (LP,UB),
is {\em reversed}, and not placed symmetrically with respect to MC1 for $d\neq 0$.

% Reflectance, transmittance and absorbance. Formation of zero GVD points
Eqs. (\ref{m1}-\ref{m4}) allows calculation of
the linear reflectivity $R_{j}$ and absorbance $A_{j}$ as a function of $\delta$ for finite gratings, see Fig. \ref{fig1n}(d). Physical dispersion branches
are selected with the criterion $R\le 1$, see also Fig. \ref{fig2n}(a). It is evident that the polariton feature introduces an {\em anomalous transmittance} region
located inside the bandgap, at a detuning $d$ from the bandgap
center, which divides the bandgap into two smaller sub-bandgaps
\cite{vardeny}. Furthermore, we observe here that the formation of such sub-gaps is accompanied by the
appearance of {\em two zero group-velocity dispersion (GVD) points}, therefore dividing the spectrum
into alternating regions of anomalous [$(\de^{2}q_{j}/\de\delta^{2})<0$] and
normal [$(\de^{2}q_{j}/\de\delta^{2})>0$] GVD [see red dots in Figs.
\ref{fig2n}(a,b)]. It is worth noticing that {\em in conventional gratings such points do not exist}, while MC polaritons possess only one
inflection point on their lower branch, which
is of paramount importance for the occurrence of the nonlinear effects
observed in experiments \cite{savvidis,kavokin}. We shall see that the
full significance of this analogy goes well beyond the mere similarity
in the dispersive properties.

% group velocity, slow light
Due to its large excitonic component and flat dispersion, the UB (LB) branch with $d<0$ ($d>0$) turns out to be a potential candidate
for slow-light-enhanced nonlinear optics at
low intensities \cite{slowlight}. This is confirmed by calculating the group velocities
(GVs) of all branches for $d=0$ and $d=-0.04$ [Fig.
\ref{fig2n}(c) and (d) respectively]. The UP/LP branches show GVs
approaching the speed of light in the medium [normalized to $\pm 1$
in Fig. \ref{fig2n}(c,d)] for large values of $\delta$, as in
conventional gratings. The GV can be reduced considerably when
$\delta$ approaches the band edges, but then progressively less photons
will be available due to evanescence of the electric field in those
regions \cite{slowlight}. However,
Figs. \ref{fig2n}(c,d) show that in proximity of the UB/LB branches
($\delta\sim d$), small GVs can be obtained. For $d=0$, Fig.
\ref{fig2n}(c), the central GV curve is symmetric, while, for
$d\neq 0$, the GV curve is asymmetric, see Fig. \ref{fig2n}(d).
Close to $\delta\simeq d$, exciton absorption can strongly
affect both UB/LB branches, depending on the precise value of
$\tilde{\gamma}_{\rm x}$. Being mostly excitonic in nature and spectrally narrower, the
UB (LB) branch is the one most affected by exciton absorption for $d<0$ ($d>0$).
Moreover, a reduction in group velocity comes at the price of a reduced bandwidth, which is a fundamental limitation
for all slow-light devices \cite{slowlight}. This
corresponds here to the progressive straightening of the flat UB branch for large values of $|d|$.
At low temperatures, typically $T< 5$ K, there are frequency regions for which absorption
is reasonably small and relatively far from the band edges, where the group velocity is
greatly reduced. For instance, Fig. \ref{fig2n}(d) shows that an increase of the group index up to $20$ is achieved
for realistic parameters corresponding to the ZnO-based grating structure discussed above.
By increasing the interaction time between the medium and light field of
approximately an order of magnitude, it is possible to achieve low-light-level nonlinear optics
\cite{slowlight}. This will open up new vistas for slow-light-enhanced nonlinearities in gratings.

% effective Braggoriton mass
We have calculated that the physical effective polariton masses at $q=0$ for each dispersive branch $m_{j}$ are given by
\eq{effectivemass1}{m_{j}=\frac{\hbar\pi\sqrt{\eps_{b}}}{2c\lambda_{\rm
x}}\mathcal{M}_{j}(d),}
where $\mathcal{M}_{j}(d)\equiv[(\de^{2}\delta_{j}/\de
q^{2})]^{-1}_{q=0}$ is a $d$-dependent coefficient shown in Fig. \ref{fig3n}(a). The
prefactor $\hbar\pi\sqrt{\eps_{b}}/(2c\lambda_{\rm x})$ assumes the value
of $\approx 4\cdot 10^{-6}m_{e}$ for bulk ZnO, which is
of the order of magnitude of the cavity photon mass in the upper
polariton branch of MCs, the latter being in practice not usable for nonlinear
optics purposes due to the absence of inflection points in the
polariton dispersion \cite{kavokin}. Thus 1D polaritonic gratings allow the possibility to
obtain very small polariton masses compared to the lower
branch of conventional MC-polaritons (the mass of which is $\sim 10^{-4}m_{e}$
\cite{kavokin}). The above considerations, valid for 1D structures, suggest that when the concept of resonant
polaritonic grating is extended to a 2D periodic system (by considering
for example structures similar to those described in Ref. \cite{gerace}), there
will be realistic chances to obtain dispersions possessing extremely small masses, thus suitable for
experiments on Bose-Einstein condensation (BEC) and superfluidity of polaritons \cite{bec}.

% parametric scattering, phase-matching at magic frequencies, matrix elements
Owing to their excitonic content, Braggoritons are very
nonlinear due to the strong interactions of their constituents, thus giving rise to
strong optical nonlinearities of Kerr (cubic) type. The nonlinear coefficient at the exciton resonance can be {\em several orders of
magnitude larger than the one for highly-nonlinear optical fibers} \cite{zhang}. We report here that a stimulated parametric scattering process occurs for Braggoriton states, which is the analogue of the parametric amplification in semiconductor MCs originally proposed and demonstrated in 2000 by Savvidis and Baumberg \cite{savvidis}. In this experiment, when an intense pump excites the lower MC polariton branch at a certain 'magic angle'
(corresponding to a magic wavenumber, located near the inflection point), one observes a large amplification of a weak probe beam which stimulates the same MC lower branch at normal incidence. 
Such an amplification is due to the scattering of the two pumped polariton states into a pair of signal and idler polaritons \cite{savvidis}, which follows from the phase-matching conditions (energy and momentum conservation). Only modulationally unstable frequencies can grow rapidly enough to produce the exponential amplification of the weak probe (located at $q=0$) at the expense of the pump.
In our system, the phase-matching condition is provided by the nonlinear terms in Eqs. (\ref{m3}-\ref{m4}), and is satisfied in close proximity of {\em two} inflection points [located in the UB and LB branches as shown in Fig. \ref{fig2n}(a,b)] - note that only {\em one} inflection point is available in MC polaritons -  thus effectively giving rise to amplification of Braggoriton states at $q=0$. As we are dealing with a grating structure, these two inflection points can be externally excited just by changing the frequencies of the input pump and probe pulses, see also Fig. \ref{fig3n}(b) and (e). Hence, Braggoriton amplification is characterized by two {\em magic frequencies} 
[$\delta_{m1,m2}$ in Fig. \ref{fig3n}(b)], in correspondence of which one transfers polaritons from the pump ($q=q_{m}$) to states with $q=0$ [see Fig. \ref{fig3n}(b)]. The final states are located symmetrically in both frequency and wavenumber space.

The scattering processes on the various dispersion branches can have very different matrix elements, and thus the efficiency can vary
considerably for different branches, as shown in Figs. \ref{fig3n}(c,d).
The intraband matrix elements of the UP and LP branches are shown in Fig. \ref{fig3n}(d). They are negligible for small values of $|d|$, as they are not phase-matched, and thus the efficiency of the process on these branches is very low.
The opposite scenario is observed for the UB/LB branches, see Fig. \ref{fig3n}(c). As a typical example, when $d=-0.04$, at a value of $|q|\approx 0.38$, where the zero GVD points are located, the ratio between the matrix element for the UB branch and the one for the LB branch is approximately $r_{UB/LB}\equiv|M_{UB}/M_{LB}|\approx 6$, so that
the UB branch is six times 'more nonlinear' than the LB branch for the specific material parameters chosen. This ratio can be tuned up by several orders of magnitude
by just increasing the value of $|d|$. For instance, for $d=-0.2$ we would have $r_{UB/LB}\approx 2\times 10^{3}$ at the zero GVD points. This allows a very large
tunability of the effective nonlinearity of the various branches in polaritonic gratings.

In conclusion, we have studied some previously unexplored
nonlinear properties of 1D Bragg gratings, when
the PBG is near-resonant with an excitonic feature of the
medium. We have developed a new model based on a set
of coupled-mode equations [Eqs. (\ref{m1}-\ref{m4})] that is able to
accurately describe the linear and nonlinear propagation
of Braggoriton states, which will be instrumental for future
theoretical investigations. The novel feature of such
gratings is the formation of two inflection points in the
dispersion characteristics, leading to the formation of 
regions of slow-light enhanced nonlinear
propagation, as well as the existence of a strong
parametric amplification at two Õmagic frequenciesÕ.
We believe that the results shown in this Letter are particularly important 
in the field of nonlinear parametric processes with exciton-polaritons. 
We are confident that the present work will stimulate further 
investigations of similar effects in PhCr structures 
of reduced dimensionality (for both exciton and photon fields) 
with potential applications to polariton quantum optics and 
polariton macroscopic coherence in mixed excitonic-photonic systems.

F.B. is supported by the German Max Planck Society for the Advancement of
Science (MPG) and by the UK Engineering and Physical Sciences Research Council (EPSRC).

\end{document}